\documentclass[letterpaper]{article}
\usepackage{graphicx}
\usepackage{hyperref}

\begin{document}

\title
{Meson Mass Spectrum of Heavy-Light Quarks Combinations with Dirac Equation}
\author{B. Pan
\footnote{email: \href{mailto:pan_b@sina.com}{pan\_b@sina.com}.}
\footnote{Member of the $BaBar$ Collaboration at slac.stanford.edu since 2004.}}
\date{\today}
\maketitle

\begin{abstract}
We use the Dirac equation to study the mass spectrum of mesons
with heavy-light quark combinations. First we study the Dirac
equation with spherically symmetry and funnel potential, and
apply them on the hydrogen-like atom problem to check the
correctness of our numerical program.
Then we test the parameters in Olsson's paper~\cite{Olsson}.
We show that Olsson's parameters are good in fitting the
averaged central mass, but fail to get correct energy
fine splitting. Finally we fit the mass spectrum data of
$D$, $D_s$, $B$ and $B_{s}$ mesons with our parameters
by solve the Dirac equation
and funnel potential, calculate the energy
splitting of the $S$ and $P$ states.
Our parameters can fit the mass and fine splitting
with errors in less than $7MeV$.
\end{abstract}

%mass models, 12.40.Yx
%properties of Mesons, 14.40.-n
%Quark models, 12.39.-x
%Dirac equation, 03.65.Pm


\section{Meson Mass Spectrum Question}

In the standard model, a meson is composed of a quark and an anti-quark,
bound together by the strong interaction.
Through the studying of the mass spectrum of mesons, we can demonstrate the
correctness of the quantum fields theory, and predict the particle's mass
that has not been found yet in the experiments.

Many articles had studied the meson mass spectrum problems and got many
very good results~\cite{Olsson}~\cite{Ebert}~\cite{Eichten}~\cite{Cornell-1}.
Most authors used the Schr$\ddot{o}$dinger equation to solve the problem.
Their treatment is accurate enough for the heavy mesons,
which are composed by two heavy quarks and moving slowly,
thus may be treated non-relativistically. Let's take an estimate.
If we think the mass of a meson is the combination of the total mass,
kinetic and potential energies of the two composition quarks,
the binding energy (kinetic + potential) is calculated and
listed in Table~\ref{table:quark}.
The constituent mass~\cite{pdg} values we used are: $m_u=0.30$ GeV,
$m_d=0.30$ GeV, $m_s=0.45$ GeV, $m_c=1.3$ GeV, $m_b=4.5$ GeV
and $m_t=180$ GeV. The "ratio" column is defined as
\begin{equation}
\textnormal{Ratio}=\frac{\textnormal{binding energy}}{\textnormal{light quark's mass}}\nonumber
\end{equation}
Compared to the light quark's mass in the $J/\psi$, $\Upsilon$
and $B_c$ mesons, their kinetic and potential energies
are not very large.
But in the $D$, $D_s$, $B$ and $B_s$ mesons,
the ratio is very large. So the light quarks in these mesons,
$D$, $D_s$, $B$ and $B_s$, are moving with relativistic energies.
Thus requires us to use relativistic equation, the Dirac equation,
to solve the spectrum problem.

\begin{table}[ht]
\centering
\caption{The composition quarks and mass of some mesons~\cite{Olsson}~\cite{Ebert}~\cite{pdg}.}
\begin{tabular}
%{|c|c|c|c|c|}
{ccccc}
\hline\hline
$\matrix{{\textnormal{Meson's}}\\{\textnormal{name}}}$ & $\matrix{{\textnormal{Meson}}\\{\textnormal{mass}}\\{(GeV)}}$ &  $\matrix{{\textnormal{Composition}}\\{\textnormal{quarks}}}$ & $\matrix{{\textnormal{Binding}}\\{\textnormal{energy}}\\{(GeV)}}$ & Ratio \\
\hline
$D$&1.975&$\matrix{{c\bar u}\\{c\bar d}}$&0.37&1.23\\
%\hline
$D_s$&2.075&$c\bar s$&0.32&0.75\\
%\hline
$B$&5.314&$\matrix{{u\bar b}\\{d\bar b}}$&0.37&1.23\\
%\hline
$B_s$&5.410&$s\bar b$&0.32&0.75\\
%\hline\hline
$J/\psi$&3.097&$c\bar c$&0.40&0.28\\
%\hline
$\Upsilon$&9.464&$b\bar b$&0.40&0.09\\
%\hline
$B_c$&6.264&$c\bar b$&0.40&0.28\\
\hline\hline
\end{tabular}
\label{table:quark}
\end{table}

We should point out that the quark's mass we used here is the
constituent quark mass, which is the quark's current mass
plus the mass of the gluon fields and sea-quarks,
as the effective quark mass of the valence quark.
The current quark mass means the mass of a quark itself only.
Values of the composition mass and current mass differ greatly.
For example, proton's mass is about 0.938 GeV.
The rest current masses of its three valence quarks are only
about 0.011 GeV each. But we can treat the mass of each up
or down quark with constituent mass as large as 0.30 GeV.

In a charm meson, there are two quarks, with $m_c\gg m_{u,d}$. So we may
simplify the problem by treat the heavier quark's mass as infinite.
Then the problem is reduced to the light quark moving in the funnel
potential that created by the heavier quark. Solving the Dirac
equation with the funnel potential, we can get the eigenenergy of
the light quark in the funnel potential. Because of the Dirac
equation includes the light quark's spin-orbit interaction, the
eigenenergy will reflect the splitting of the $S$ and $P$ states.
The influence of the heavier quark is treated by
adding its mass in the spin dependant force. The total mass of the
meson can be expressed as:
\begin{equation}
M_{q_1 \bar q_2} = m_{q_1} + m_{\bar q_2} + E + \Delta,
\label{eq:mqq}
\end{equation}
that's the sum of the masses of the two quarks $m_{q_1}$ and
$m_{\bar q_2}$, eigenenergy of the system $E$,
and the hyper-fine energy splitting $\Delta$.

M. G. Olsson~\cite{Olsson} shows that the Dirac equation can
be used to get a good fitting in the average mass spectrum.
But he did not calculate the fine structure splitting.

D. Ebert, V. O. Galkin and R. N. Faustov~\cite{Ebert} use
Schr$\ddot{o}$dinger equation with relativistic potential
to solve the mass spectrum and calculate the fine splitting.
They point out that in the spin dependent potentials, we should use
$1\over{eigenenergy}$, instead of use $1\over {m_q}$ in
Eichten and Quigg's paper \cite{Eichten} to get energy splitting.

In this paper, we will solve the Dirac equation numerically to fit
the meson's mass spectrum of the heavy-light quarks combination system,
such as $D$, $D_s$, $B$ and $B_s$, and calculate the spin-orbit
energy hyper fine splitting. Our result will be compared to
the experimental data.

\section{Dirac Equation with Spherically Symmetry}

The Dirac equation for free particle is
\begin{equation}
(\gamma _\mu {\partial  \over {\partial x_\mu }}+m)\psi =0.
\label{eq:dirac}
\end{equation}

If the potential has spherical symmetry, and can be written as
a combination of Lorentz scalar part $V_s(r)$ and vector
Coulomb potential $V_v(r)$, then the Dirac equation can
be written as~\cite{Bjorken}~\cite{Groos}~\cite{Greiner1}:
\begin{equation}
i{{\partial \psi } \over {\partial t}}=\left[ {-i\alpha_{i}{\partial  \over {\partial x_i}}+\beta \big( m + V_s(r) \big ) + V_v(r)} \right]\psi,
\label{eq:dirv}
\end{equation}
with the solution has the form like
\begin{equation}
\psi (r)=
\left( {\begin{array}{*{20}c}
   F(r)  \\
   G(r)  \\
\end{array}} \right)
= \left( {\begin{array}{*{20}c}
   f^\pm (r)y_{jj_z}^\pm (\hat r)  \\
   ig^\pm(r)y_{jj_z}^\mp (\hat r)  \\
\end{array}} \right),
\label{eq:fg}
\end{equation}
in which $y_{jj_z}$ is a two component generalization of the spherical
harmonic functions $Y_{l\,j_z}$,
\begin{equation}
y_{jj_z}^k(\hat r)=-{k \over {|k|}}\sqrt {{{k+{1 \over 2}-j_z} \over {2k+1}}}\alpha Y_{l,j_z -{1 \over 2}}(\hat r)+\sqrt {{{k+{1 \over 2}+j_z} \over {2k+1}}}\beta Y_{l,j_z +{1 \over 2}}(\hat r).
\end{equation}
The number $k$ is
\begin{equation}
k=\pm \left( {j+ \frac {1}{2}} \right),
\end{equation}
in which the $\pm$ sign is
\begin{equation}
 \left\{ {\begin{array}{*{20}l}
   \mbox{if} & l=j+\frac {1}{2} &\Rightarrow &k=\left( {j+ \frac {1}{2}} \right)=l,  \\
   \mbox{if} & l=j-\frac {1}{2} &\Rightarrow &k=-\left( {j+ \frac {1}{2}} \right)=-(l+1).  \\
\end{array}} \right.
\end{equation}
Thus the coupled equations for the radial functions are
\begin{eqnarray}
\Big [ E - m - V_s(r)- V_v(r) \Big ] f(r) & = & -\frac{dg(r)}{dr}- \frac{1-k}{r}g(r),
\label{eq:ffgg1}
\end{eqnarray}
\begin{eqnarray}
\Big [ E + m + V_s(r) - V_v(r) \Big ] g(r) & = & \frac{df(r)}{dr}+ \frac{1+k}{r}f(r).
\label{eq:ffgg2}
\end{eqnarray}

Notes that the angular dependent variables have been removed,
and only two radial coordinate dependent
functions $f$ and $g$ are left in the equation. By solving the
equations (\ref{eq:ffgg1}) and (\ref{eq:ffgg2}), we will get
the energies and eigenfunctions of a relativistic particle
that moves in the central potential.

\section{Potential inside Meson}

\subsection{QCD One Gluon Exchange Coulomb-like Potential}

Quantum Chromodynamics (QCD) describes the strong interaction
among quarks and gluons.
The Lagrangian density in QCD can be written as:
\begin{equation}
{\cal L} = -{1 \over 4}F_{\mu \nu }^aF_a^{\mu \nu }+\sum\limits_{f=1}^{n_F} {\bar q_f(i\not D-m_f)q_f},
\end{equation}
in which
\begin{eqnarray}
F_{\mu \nu }^a & = & \partial_\mu A_\nu^a - \partial_\nu A_\mu^a + g_s \varepsilon_{abc} A_\mu ^aA_\nu ^b, \nonumber \\
 \not D_\mu & = & \partial_\mu -i g_s T_a A_\mu ^a, \nonumber
\end{eqnarray}
where $q_f$ is the quark field, $g_s$ is the strong interaction
coupling strength, $T_a$ are the generators of
the color $SU(3)$ group, $\varepsilon_{abc}$ is the structure
constant of the $su(3)$ Lie algebra, and $A_\mu$ is the $SU(3)$ color
gauge field.

Effective potential is used to study the bound states of the
quarks in mesons. The one gluon exchange Coulomb-like
potential~\cite{Griffiths} between two quarks is
\begin{equation}
V_v(r)=-\frac{4}{3}\frac{\alpha _s}{r},
\end{equation}
in which
\begin{equation}
\alpha _s={{g_s^2} \over {4\pi }}.
\end{equation}

\subsection{Funnel Potential}

At large distances, there should exist a confine potential that
describes the color confinement. Unfortunately, till now,
the color confinement can not be derived from the QCD first principle.
So we may add it in "by hand". The commonly used Cornell confinement
potential~\cite{Cornell-1}~\cite{Cornell-2}~\cite{Cornell-3} is
\begin{equation}
V_s(r) = V_{confine}(r)=ar,
\end{equation}
which will make the combination of the scalar $V_s(r)$ and vector $V_v(r)$
potential has a funnel shape. The spectrum obtained with the
funnel potential is in good agreement with experimental
data for the light and heavy mesons~\cite{Olsson}~\cite{Ebert}~\cite{Eichten}.
Typical values for the parameters are $a \approx 0.2 \;GeV^2$, and
$\alpha_s \approx 0.2 \sim 0.3$.

\section{Spin Dependant Hyper-fine Energy Splitting}

In QED, the relativistic Dirac equation describes two or more massive
spin 1/2 particles interacting electromagnetically. The perturbation QED formula of the fine splitting includes spin-orbit interaction ($L\cdot
S$), spin-spin interaction ($s_i \cdot s_j$), and Breit spin-spin tensor
interaction ($s_i \cdot s_j - \frac {(s_i \cdot r_{ij})(s_j \cdot
r_{ij})}{r_{ij}^2}$) terms.

Within the framework of QCD, there are similar potentials.
If the Schr\"{o}dinger equation is used to solve the mass spectrum
problem, the spin-dependent hyper-fine energy splitting to the
first order of $\alpha_s$ can be written as~\cite{Eichten}:
\begin{equation}
 \Delta = \sum\limits_{k = 1}^4 {T_k }
\label{eq:delta}
\end{equation}
with
\begin{eqnarray}
 T_1  & = & \frac{{\langle L \cdot s_i \rangle }}{{2m_i^2 }}\tilde T_1 (m_i ,m_j ) + \frac{{\langle L \cdot s_j \rangle }}{{2m_j^2 }}\tilde T_1 (m_j ,m_i ), \nonumber \\
 T_2  & = & \frac{{\langle L \cdot s_i \rangle }}{{m_i m_j }}\tilde T_2 (m_i ,m_j ) + \frac{{\langle L \cdot s_j \rangle }}{{m_i m_j }}\tilde T_2 (m_j ,m_i ), \nonumber \\
 T_3  & = & \frac{{\langle s_i  \cdot s_j \rangle }}{{m_i m_j }}\tilde T_3 (m_i ,m_j ), \nonumber \\
 T_4  & = & \frac{{\langle S_{ij} \rangle }}{{m_i m_j }}\tilde T_4 (m_i ,m_j ),
\label{eq:t3}
\end{eqnarray}
and
\begin{eqnarray}
 \tilde T_1 (m_i ,m_j ) & = &  - \langle \frac{1}{r}\frac{{dV}}{{dr}}\rangle  + 2\tilde T_2 (m_i ,m_j ), \nonumber \\
 \tilde T_2 (m_i ,m_j ) & = & \frac{{4\alpha _s }}{3}\langle r^{ - 3} \rangle,  \nonumber \\
 \tilde T_3 (m_i ,m_j ) & = & \frac{{32\pi \alpha _s }}{9}|\psi (0)|^2,  \nonumber \\
 \tilde T_4 (m_i ,m_j ) & = & \frac{{\alpha _s }}{3}\langle r^{ - 3} \rangle ,
\label{eq:tt3}
\end{eqnarray}
in which
\begin{equation}
 S_{ij} = 4[3(s_i  \cdot \hat n)(s_j  \cdot \hat n) - s_i  \cdot s_j ].
\end{equation}

Now we use the Dirac equation to solve the mass spectrum.
The energy splitting terms in equ.(\ref{eq:t3}) and (\ref{eq:tt3})
should be modified a little bit.
In the non-relativistic limit, $p=mv<<mc$.
If we let $V_s=0$ and $V=V_v$, the Dirac equation (\ref{eq:dirv})
and wave function (\ref{eq:fg}) will be reduced to:
\begin{equation}
\left [ \frac{\mbox{\boldmath$\hat{p}^2$}}{2m}+V \right ]F(r)-
\frac{\hbar ^2}{4m^2c^2}\frac{dV}{dr}\frac{dF}{dr} +
\frac{\hbar}{2m^2c^2}\frac{1}{r}\frac{dV}{dr}\mbox{\boldmath$\hat{s}$}
\cdot \mbox{\boldmath$\hat{L}$}\;F(r)= (E - mc^2) \cdot F(r).
\label{eq:dirac-reduced}
\end{equation}

Compare to the Schr\"{o}dinger equation, the non-relativistic limited Dirac
equation(\ref{eq:dirac-reduced}) has already included the $\frac{1}{2m^2
r}\frac{dV}{dr}\mbox{\boldmath$\hat{s}$} \cdot
\mbox{\boldmath$\hat{L}$}$ spin-orbit interaction term.
So when we use the Dirac
equation to study energy splitting, the $-\langle \frac{1}{r}\frac{{dV}}{{dr}}\rangle$ term should be removed from
$\tilde T_1 (m_i ,m_j )$ in equ.(\ref{eq:tt3}).
Since we treat the heavier quark's mass as infinite in
the Dirac equation, the fracture $1\over{m_i}$ will be
replaced by a meaningful limited number: over energy eigenvalue,
$1\over E$.
Now the terms in the energy splitting equ.(\ref{eq:delta})
will be
\begin{eqnarray}
T_1&=& {{\langle L\cdot s_i\rangle } \over {2E^2}}\tilde T_1(m_i,m_j) + {{\langle L\cdot s_j\rangle } \over {2Em_j}}\tilde T_1(m_j,m_i), \nonumber \\
T_2&=&{{\langle L\cdot s_i\rangle } \over {Em_j}}\tilde T_2+{{\langle L\cdot s_j\rangle } \over {Em_j}}\tilde T_2, \nonumber \\
T_3&=&{{\langle s_i\cdot s_j\rangle } \over {Em_j}}\tilde T_3, \nonumber \\
T_4&=&{{\langle S_{ij}\rangle } \over {Em_j}}\tilde T_4,
\label{eq:t4}
\end{eqnarray}
and
\begin{eqnarray}
 \tilde T_1 (m_i ,m_j ) & = & 2\tilde T_2 (m_i ,m_j ), \nonumber \\
 \tilde T_2 (m_i ,m_j ) & = & \frac{{4\alpha _s }}{3}\langle r^{ - 3} \rangle,  \nonumber \\
 \tilde T_3 (m_i ,m_j ) & = & \frac{{32\pi \alpha _s }}{9}|\psi (0)|^2,  \nonumber \\
 \tilde T_4 (m_i ,m_j ) & = & \frac{{\alpha _s }}{3}\langle r^{ - 3} \rangle .
\label{eq:tt4}
\end{eqnarray}

There is a trick in calculating the ${\psi (0)}^2$ term in equ.(\ref{eq:tt4}).
It can be replaced by the production of the wave
function and a $\delta$ function: $\delta (r){\psi (r)}^2$. The
$\delta$ function can be defined as
\begin{equation}
\delta (r)\to \mathop {\lim }\limits_{b\to 0}{1 \over {b^3\pi ^{{3
\over 2}}}}e^{-{{r^2} \over {b^2}}}.
\end{equation}
So
\begin{eqnarray}
\tilde T_3(m_i,m_j)&=&{{32\pi \alpha _s} \over 9}|\psi (0)|^2\nonumber\\
&=&{{32\pi \alpha _s} \over 9}\int_0^\infty  {\psi ^2(r)\delta (r)r^2dr}\nonumber\\
&=&\mathop {\lim }\limits_{b\to 0}{{32\pi \alpha _s} \over {9b^3\pi
^{{3 \over 2}}}}\int_0^\infty  {\psi ^2(r)}e^{-{{r^2} \over
{b^2}}}r^2dr,
\end{eqnarray}
which can be easily integrated numerically.

By adding the spin dependant part (\ref{eq:delta}) with (\ref{eq:t4})
and (\ref{eq:tt4}), into the solution of equ.(\ref{eq:ffgg1})
and equ.(\ref{eq:ffgg2}), we will get the meson spectrum with
fine structure energy splitting.

\section{Numerical Results for the Mesons}

We will use the double shooting method and Runge-Kutta 4th method to solve the Dirac equation.
Because the Dirac equation for the hydrogen-like atom has exact analytical solution,
we first run our program on the hydrogen-like atom case to test whether our code works or not.
The test results are shown in Appendix A, that we can get up to $10^{-5}$ accuracy.
Next we run with Olsson's parameters and the funnel potential to compare to
Olsson's~\cite{Olsson} results, which are shown in Appendix B.

Finally with the funnel potential
\begin{eqnarray}
V_s(r) &=& ar, \\
V_v(r) &=& - \frac {\kappa}{r},
\end{eqnarray}
we solve the Dirac equation numerically with our own parameters.
Because meson has a confining part of potential,
the wave function tends to contract to the center.
So the system will be smaller in scale than the hydrogen-like atom,
which does not have a confining part. According to our test,
we choose the boundary condition as
\begin{equation}
\psi (r)|_{r=0\:(GeV^{-1})}=0,
\end{equation}
and
\begin{equation}
\psi (r)|_{r\ge 20\:(GeV^{-1})}=0.
\end{equation}
The radial part of the wave functions will be normalized with the
Simpson's integration rule to one,
\begin{equation}
\int_{0}^{\infty} (f^2 + g^2) r^2dr = 1.
\end{equation}

The fitting method is done by
inputting two mesons masses as initial values to determine the
parameters in the expression (\ref{eq:delta}), equ (\ref{eq:ffgg1}) and (\ref{eq:ffgg2}).
Then use the trial parameters, input configuration parameters
of an unknown meson, to get the mass of that unknown meson.

After many trials, we find out the following set of parameters can fit the meson's average mass and splitting very well.
\begin{eqnarray}
m_{ud} &=& 0.280\;GeV, \nonumber\\
m_s &=& 0.429\;GeV, \nonumber\\
m_c &=& 1.095\;GeV, \nonumber\\
m_b &=& 4.435\;GeV, \nonumber\\
a &=& 0.368\;GeV^2, \nonumber\\
\kappa &=& 0.400.
\label{eq:para}
\end{eqnarray}

\begin{table}[htb]
\caption{Using Dirac equation to fit the meson spectrum.
Particle's experimental mass are from the PDG~\cite{pdg} book.
Mass values are in $MeV$.}
\centering
\begin{tabular}
%{|c|c|c|c|c|c|c|}
{ccccccc}
\hline\hline
 & & Spin & & Numerical & Numerical & \\
States & ${}^{2S+1}L_J$ & averaged  & $k$ & center & splitting & Parameter \\
 & & mass & & mass & mass & b \\
\hline

${\begin{array}{*{20}c} D(1867)\\ D^*(2010) \\ \end{array}}$ &
${\begin{array}{*{20}c} {}^{1}S_0 \\ {}^{3}S_1 \\ \end{array}}$ & 1S(1974) & -1 & 1975 & ${\begin{array}{*{20}c} 1867\\ 2010 \\ \end{array}}$ & b=1.07\\
%&&&&&&\\
%\hline
&&&&&&\\

${\begin{array}{*{20}c} D_{1}(2423)\\ D^{*}_2(2457) \\ \end{array}}$ &
${\begin{array}{*{20}c} {}^{1}P_1 \\ {}^{3}P_2 \\ \end{array}}$ & 2P(2444) & -2 & 2407 & ${\begin{array}{*{20}c} 2423\\ 2457 \\ \end{array}}$ & N/A\\
&&&&&&\\

${\begin{array}{*{20}c} D_{s}(1969)\\ D^{*}_s(2110) \\ \end{array}}$ &
${\begin{array}{*{20}c} {}^{1}S_0 \\ {}^{3}S_1 \\ \end{array}}$ & 1S(2075) & -1 & 2074 & ${\begin{array}{*{20}c} 1968\\ 2109 \\ \end{array}}$ & b=1.08\\
&&&&&&\\

${\begin{array}{*{20}c} D_{s1}(2535)\\ D_{sJ}(2573) \\ \end{array}}$ &
${\begin{array}{*{20}c} {}^{1}P_1 \\ {}^{3}P_2 \\ \end{array}}$ & 2P(2559) & -2 & 2515 & ${\begin{array}{*{20}c} 2528\\ 2570 \\ \end{array}}$ & N/A\\
&&&&&&\\

${\begin{array}{*{20}c} B(5279)\\ B^{*}(5325) \\ \end{array}}$ &
${\begin{array}{*{20}c} {}^{1}S_0 \\ {}^{3}S_1 \\ \end{array}}$ & 1S(5314) & -1 & 5314 & ${\begin{array}{*{20}c} 5279\\ 5325 \\ \end{array}}$ & b=0.87\\
&&&&&&\\

${\begin{array}{*{20}c} B_{s}(5375)\\ B^{*}_s(5421) \\ \end{array}}$ &
${\begin{array}{*{20}c} {}^{1}S_0 \\ {}^{3}S_1 \\ \end{array}}$ & 1S(5410) & -1 & 5412 & ${\begin{array}{*{20}c} 5376\\ 5422 \\ \end{array}}$ & b=0.88\\
\hline\hline
\end{tabular}
\label{meson-fit}
\end{table}

Our fitting results are listed in Table~\ref{meson-fit},
in which the particle's experimental mass are from the PDG~\cite{pdg} book.
Spin averaged mass is calculated by taking $3 \over 4$ ($5 \over 8$)
of the triplet mass, and $1 \over 4$ ($3 \over 8$) of the
singlet mass for the s(p) states~\cite{Olsson}.
The column "Numerical center mass" are the numerical result of the central
mass of the $S$ and $P$ states. Then we use the fine structure
formula (\ref{eq:delta}) to calculate the energy fine splitting that
are listed in the column "Numerical splitting mass".
For the $P$ states, by intentionally choosing parameters that let the spin
average mass does not sit between the $^{1}P_1$ and ${}^{3}P_2$
states, but let the average mass lower than
both of the $^{1}P_1$ and ${}^{3}P_2$
states, we can get good fittings for their splitting.
The errors for $S$ states are about $1\;MeV$, while
the $P$ states errors are less than $7\,MeV$.

We also calculate the average values of $r$, $r^2$, $r^{-1}$ and $r^{-2}$, which are listed in Table~\ref{meson-expect}. In our model, the wave functions are related to the light quark's mass, but not to the heavy quark's mass.

\begin{table}[htb]
\caption{Average value of $r$, $r^2$, $r^{-1}$ and $r^{-2}$, with values are in $GeV^{n}$.}
\centering
\begin{tabular}
%{|c|c|c|c|c|c|c|}
{ccrrrr}
\hline\hline
&&&&&\\
Light quark & ${}^{2S+1}L_J$ & $<r>$  & $<r^2>$ & $<r^{-1}>$ & $<r^{-2}>$ \\
\hline

u/d & ${\begin{array}{*{20}c} {}^{1}S_0 \\ {}^{3}S_1 \\ \end{array}}$ & 1.520 & 2.811 & 0.887 & 1.412\\
%\hline
&&&&&\\

u/d & ${\begin{array}{*{20}c} {}^{1}P_1 \\ {}^{3}P_2 \\ \end{array}}$ & 2.205 & 5.423 & 0.521 & 0.326\\
&&&&&\\

s & ${\begin{array}{*{20}c} {}^{1}S_0 \\ {}^{3}S_1 \\ \end{array}}$ & 1.437 & 2.521 & 0.944 & 1.617\\
&&&&&\\

s & ${\begin{array}{*{20}c} {}^{1}P_1 \\ {}^{3}P_2 \\ \end{array}}$ & 2.120 & 5.026 & 0.543 & 0.355\\

\hline\hline
\end{tabular}
\label{meson-expect}
\end{table}

\begin{figure}[htb]
\centering
\includegraphics[width=0.9\textwidth]{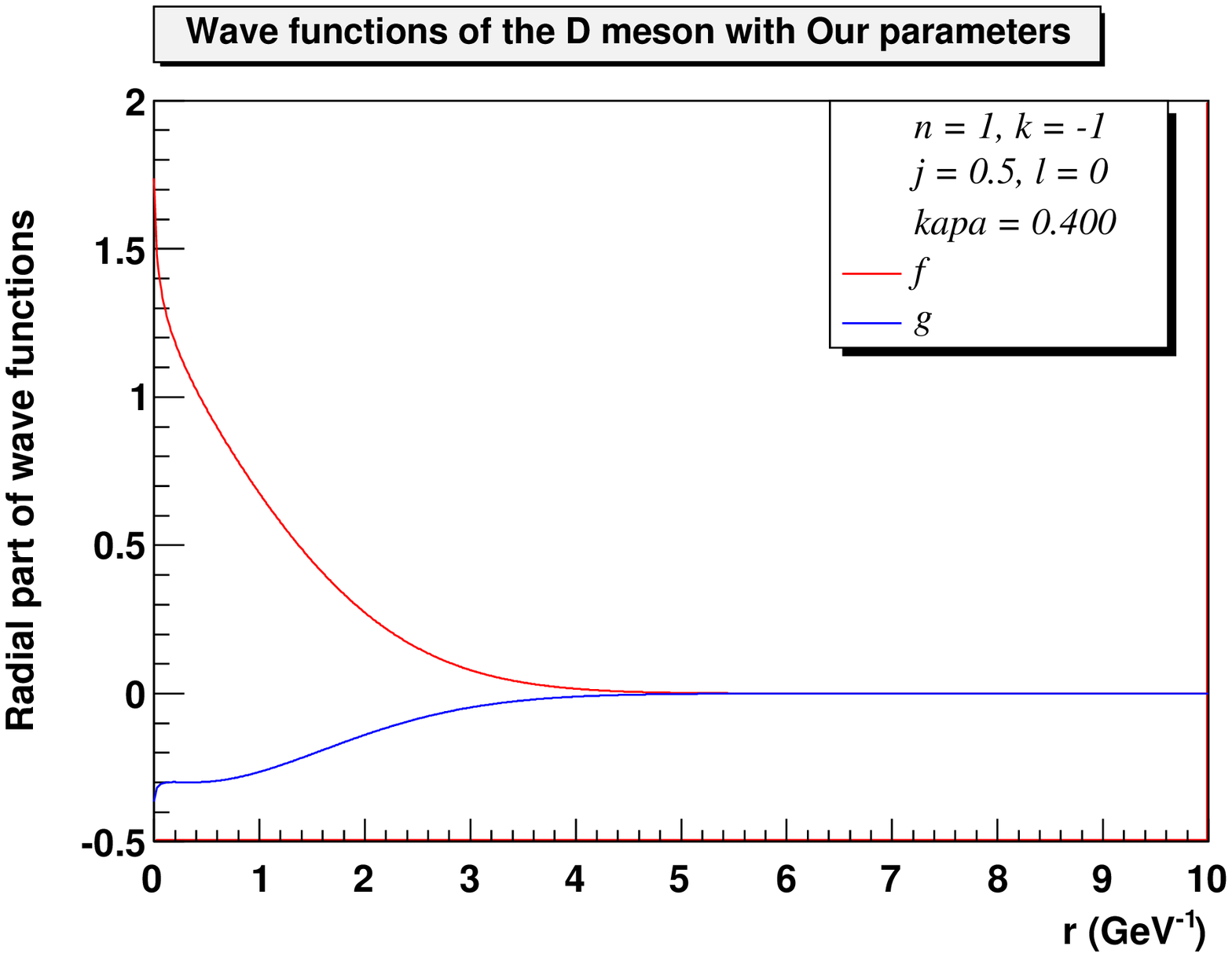}
\caption{Wave functions of the $1S$ state with our parameters.}
\label{my1}
\end{figure}

\begin{figure}[htb]
\centering
\includegraphics[width=0.9\textwidth]{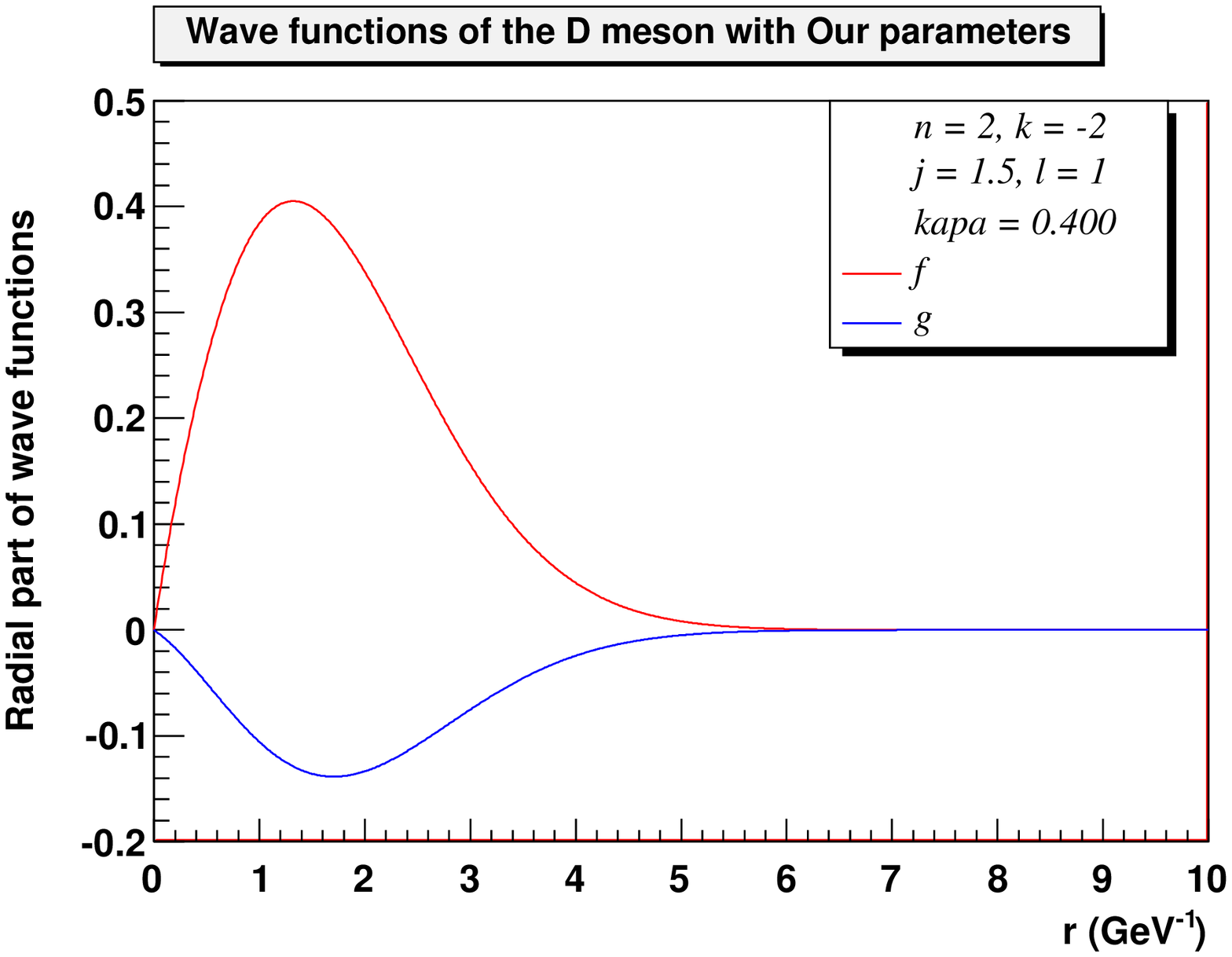}
\caption{Wave functions of the $2P$ state with our parameters.}
\label{my2}
\end{figure}

\section{Discussion}

By using our set of parameters (\ref{eq:para}), we can fit the $D$, $D_s$, $B$ and $B_s$
mesons spectrum with the errors within $\sim 7\;MeV$. The parameters
for the mass of the $u$, $d$, $s$, $c$ and $b$ quarks, $\kappa$ are
all in reasonable range. We use $a=0.368\;GeV^2$.
But in other people's paper~\cite{Ebert}~\cite{Eichten}, $a \approx
0.18\sim 0.20\;GeV^2$, in which they used with the Schr\"{o}dinger
equation. The reason may be explained as the difference between the Dirac equation and the Schr\"{o}dinger equation.
The one-gluon exchange process plus the confining potential is
\begin{equation}
V(p,q)=\sqrt {{m \over {E_p}}}\bar u(p)\left[ {-{{4\alpha _s} \over 3}{{4\pi } \over {k^2}}\gamma ^0+V_{conf}} \right]\sqrt {{m \over {E_q}}}u(q),
\end{equation}
in which
\begin{equation}
u(p)=\sqrt {{{{E_p}+m} \over {2m}}}
\left( {\begin{array}{*{20}c}
   1 \\
   {\sigma \cdot p} \over {E_p+m}  \\
\end{array}} \right),
\chi
\end{equation}
and
\begin{equation}
u^+u=\sqrt {{E \over m}}.
\end{equation}

In the non-relativistic limit, assume the exchanging gluon's energy
is small, then $E_p\approx E_q$. By using $E^2=p^2+m^2$, we can get
\begin{equation}
V\to -{{4\alpha _s} \over 3}{{4\pi } \over {k^2}}+{m \over E}V_{conf}.
\end{equation}
That means when we use confining potential $V_{conf}$ in the
Dirac equation, it is equivalent to the confining potential ${m
\over E}V_{conf}$ in the Schr\"{o}dinger equation,
\begin{equation}
V_{conf}={a r} \, \longrightarrow \, {m \over E}V_{conf}={m \over E}{a r}.
\end{equation}
So the relation between the parameters of "$\bf {a}$" in the Dirac equation and the Schr\"{o}dinger equation is
\begin{equation}
\textnormal{Dirac: }a \Longleftrightarrow \textnormal{Schr\"{o}dinger: }{m \over E}a.
\end{equation}

Let's take an estimate. For the $D$ meson, using our parameters (\ref{eq:para}), $M_{average}=1.975$ GeV, $m_u=0.28$ GeV, $m_c =1.095$ GeV, so the eigenenergy is
\begin{eqnarray}
E &=& M_{average} -m_c -m_u \nonumber\\
 &=& 1.975 - 1.095 - 0.28  \nonumber\\
 &=& 0.60 \;GeV.
\end{eqnarray}
That means the parameters of "$\bf {a}$" in the Schr\"{o}dinger equation is
\begin{equation}
{m \over E} a ={0.3 \over 0.60}\times 0.368 =0.184\;(GeV)^2,
\end{equation}
which is in the range that people used with the Schr\"{o}dinger equation.

\section*{Acknowledgments}
This project was done when I did research in the
Bowling Green State University, Ohio, U.S.A.
Liews Fulcher provided the double shooting method algorithm Fortran source code,
in which he used before with the Schr\"{o}dinger equation;
and made many helpful discussions between us.
I modified his Fortran code and did all the calculation with the Dirac equation.
Finally I got the set of parameters(\ref{eq:para}) and results(Table~\ref{meson-fit}).

%\afterpage{\clearpage}

\section*{Appendix A: Numerical Results for the Hydrogen-like Atoms}

The hydrogen-like atom is defined as a particle moves in the central Coulomb potential.
We can use its analytical solution results to test the correctness of our numerical program.

For the hydrogen-like atom with a Coulomb central potential
\begin{equation}
V=-{\kappa \over r},
\end{equation}
the exact Dirac solution~\cite{Groos}\cite{Greiner1} is
\begin{eqnarray}
E_{n,k} &=& m\left[ 1-\frac{\kappa^2}{(N+|k|)^2+2N(\sqrt {k^2-{\kappa}^2}-|k|)} \right]^{\frac{1}{2}}\nonumber\\
&=& m\left[ 1-\frac{\kappa^2}{n^2+2(n-(j+\frac{1}{2}))\left[ \sqrt {(j+\frac{1}{2})^2-{\kappa}^2}-(j+\frac{1}{2}) \right]} \right]^{\frac{1}{2}},
\label{eq:dirsol}
\end{eqnarray}
where
\begin{equation}
n=N+|k|\;\geq 1, \;\;\;\;\;\;\;\;\;\;\;\;\;\;-n\leq k<n.
\end{equation}

We use the following parameters in our numerical code.
\begin{eqnarray}
m &=& 0.3 \, GeV,\nonumber\\
Z\alpha  &=& \kappa =0.579,\nonumber\\
V &=& -{\kappa  \over r}.
\end{eqnarray}

Because the hydrogen-like atom does not have a confining part of potential,
the wave functions tend to extend far away from the center.
So the system will be large in scale. In our numerical program,
we should set a large region to solve the problem.
According to our test, we choose the boundary condition as
\begin{equation}
\psi (r)|_{r=0\:(GeV^{-1})}=0,
\end{equation}
and
\begin{equation}
\psi (r)|_{r\ge 200\:(GeV^{-1})}=0.
\end{equation}

Both of the numerical and analytical results are listed in Table~\ref{hydron-result}.
It shows that our numerical program works very well,
that there are no differences among the numerical and analytical
results of the eigenenergy to the precision of $10^{-5}$.

We also calculate the average value of $r$, $r^2$, $r^{-1}$ and $r^{-2}$,
which are listed in Table~\ref{hydron-r}, \ref{hydron-rn}
and \ref{hydron-rn2}. Our numerical results agree with the
exact Dirac solution's expect values. The Schr\"{o}dinger exact
solution's expect values are also listed for comparison.
In the non-relativistic limit, Dirac equation's $f$ part wave
function will reduce to the Schr\"{o}dinger wave function.
The difference in our results between the Dirac and
Schr\"{o}dinger equations indicate that for a system
with quark's mass and interaction strength,
we should use relativistic Dirac equation, instead of
the non-relativistic Schr\"{o}dinger equation.

\begin{table}[htb]
\small
\centering
\caption{Exact Dirac solutions and our numerical results of hydrogen-like atom.}
\begin{tabular}
%{|c|c|c|c|c|c|c|}
{ccccccc}
\hline\hline
$n$ & $k$ & $j$ & $l$ & name & Analytical energy ($GeV$) & $\matrix{{\textnormal{Numerical}}\\{\textnormal{result}}\\{(GeV)}}$ \\
\hline
&&&&&&\\
1 & -1& $1\over 2$ & 0 & $1S_{1/2}$ & $m\sqrt{1-{Za}^2}$=0.24460 & 0.24460\\
&&&&&&\\  \hline
&&&&&&\\
2 & -2& $3\over 2$ & 1 & $2P_{3/2}$ & $m\sqrt {1-{1 \over 4}(Za)^2}=0.28715$ & 0.28715\\
&&&&&&\\
{}& $\matrix{{1}\\{}\\-1}$
& $\matrix{{1\over 2}\\{}\\{1\over 2}}$
& $\matrix{{1}\\{}\\{0}}$
& $\matrix{{2P_{1/2}}\\{}\\{2S_{1/2}}}$
&  $\left. {\matrix{{}\\{}\\{}}} \right\} {m\sqrt {1-{{(Za)^2} \over {2+2\sqrt {1-(Za)^2}}}}}=0.28581$
& $\matrix{{}\\{0.28581}\\{}}$ \\
&&&&&&\\  \hline
&&&&&&\\
3 & -3& $5\over 2$ & 2 & $3D_{5/2}$ & $m\sqrt {1-{1 \over 9}(Za)^2}=0.29436$ & 0.29436\\
&&&&&&\\
{}& $\matrix{{2}\\{}\\-2}$
& $\matrix{{3\over 2}\\{}\\{3\over 2}}$
& $\matrix{{2}\\{}\\{1}}$
& $\matrix{{3D_{3/2}}\\{}\\{3P_{3/2}}}$
&  $\left. {\matrix{{}\\{}\\{}}} \right\} {m\sqrt {1-{{(Za)^2} \over {5+2\sqrt {4-(Za)^2}}}}}=0.29425$
& $\matrix{{}\\{0.29425}\\{}}$ \\
&&&&&&\\
{}& $\matrix{{1}\\{}\\-1}$
& $\matrix{{1\over 2}\\{}\\{1\over 2}}$
& $\matrix{{1}\\{}\\{0}}$
& $\matrix{{3P_{1/2}}\\{}\\{3S_{1/2}}}$
&  $\left. {\matrix{{}\\{}\\{}}} \right\} {m\sqrt {1-{{(Za)^2} \over {5+4\sqrt {1-(Za)^2}}}}}=0.29385$
& $\matrix{{}\\{0.29385}\\{}}$ \\
&&&&&&\\
\hline\hline
\end{tabular}
\label{hydron-result}
\end{table}

\begin{table}[htb]
\small
\centering
\caption{Average value of $r$ and $r^2$ of the hydrogen-like atom.
Exact Dirac (column "Exact(D)") and Schr\"{o}dinger (column "Exact(S)")
solutions are also listed for comparison to our numerical results.}
\begin{tabular}
%{|c|c|c|c|c|c|c|}
{cccc|rrr|rrr}
\hline\hline
%&&&&&\\
&&&&\multicolumn{3}{c}{$<r> (GeV^{-1})$}&\multicolumn{3}{c}{$<r^2> (GeV^{-2})$}\\
%\cline{5-10}
%\cline{5-6}

n & k & j & l & Numeric & Exact(D) & Exact(S) & Numeric & Exact(D) & Exact(S) \\
\hline
&&&&&&&&&\\
1 & -1 & $1\over 2$ & 0 & 7.572 & 7.572 & 8.635 & 79.14 & 79.14 & 99.43\\
&&&&&&&&&\\
2 & -2 & $3\over 2$ & 1 & 27.80 & 27.80 & 28.79 & 932.9 & 932.8 & 994.3\\
&&&&&&&&&\\
2 & 1 & $1\over 2$ & 1 & 24.24 & 24.25 & 28.79 & 729.7 & 730.4 & 994.3\\
&&&&&&&&&\\
2 & -1 & $1\over 2$ & 0 & 29.99 & 30.01 & 34.54 & 1074. & 1074. & 1392.\\
&&&&&&&&&\\
3 & -3 & $5\over 2$ & 2 & 59.44 & 59.47 & 60.45 & 4043. & 4050. & 4176.\\
&&&&&&&&&\\
3 & 2 & $3\over 2$ & 2 & 57.71 & 57.73 & 60.45 & 3832. & 3836. & 4176.\\
&&&&&&&&&\\
3 & -2 & $3\over 2$ & 1 & 69.16 & 69.24 & 71.96 & 5540. & 5558. & 5965.\\
&&&&&&&&&\\
3 & 1 & $1\over 2$ & 1 & 64.14 & 64.18 & 71.96 & 4798. & 4806. & 5966.\\
&&&&&&&&&\\
3 & -1 & $1\over 2$ & 0 & 69.87 & 69.94 & 77.72 & 5597. & 5611. & 6861.\\
&&&&&&&&&\\
\hline\hline
\end{tabular}
\label{hydron-r}
\end{table}

\begin{table}[htb]
\centering
\caption{Average values of $r^{-1}$ in GeV of the hydrogen-like atom.
Exact Dirac (column "Exact(D)") and Schr\"{o}dinger (column "Exact(S)")
solutions are also listed for comparison to our numerical results.}
\begin{tabular}
%{|c|c|c|c|c|c|c|}
{ccccrrr}
\hline\hline

n & k & j & l & Numeric & Exact(D) & Exact(S) \\
\hline
&&&&&&\\
1 & -1 & $1\over 2$ & 0 & $2.13\times 10^{-1}$ & $2.13\times 10^{-1}$ & $1.74\times 10^{-1}$ \\
&&&&&&\\
2 & -2 & $3\over 2$ & 1 & $4.54\times 10^{-2}$ & $4.54\times 10^{-2}$ & $4.34\times 10^{-2}$\\
&&&&&&\\
2 & 1 & $1\over 2$ & 1 & $5.59\times 10^{-2}$ & $5.59\times 10^{-2}$ & $4.34\times 10^{-2}$\\
&&&&&&\\
2 & -1 & $1\over 2$ & 0 & $5.59\times 10^{-2}$ & $5.59\times 10^{-2}$ & $4.34\times 10^{-2}$\\
&&&&&&\\
3 & -3 & $5\over 2$ & 2 & $1.97\times 10^{-2}$ & $1.97\times 10^{-2}$ & $1.93\times 10^{-2}$\\
&&&&&&\\
3 & 2 & $3\over 2$ & 2 & $2.05\times 10^{-2}$ & $2.05\times 10^{-2}$ & $1.93\times 10^{-2}$\\
&&&&&&\\
3 & -2 & $3\over 2$ & 1 & $2.05\times 10^{-2}$ & $2.05\times 10^{-2}$ & $1.93\times 10^{-2}$\\
&&&&&&\\
3 & 1 & $1\over 2$ & 1 & $2.36\times 10^{-2}$ & $2.36\times 10^{-2}$ & $1.93\times 10^{-2}$\\
&&&&&&\\
3 & -1 & $1\over 2$ & 0 & $2.36\times 10^{-2}$ & $2.36\times 10^{-2}$ & $1.93\times 10^{-2}$\\
&&&&&&\\
\hline\hline
\end{tabular}
\label{hydron-rn}
\end{table}

\begin{table}[htb]
\centering
\caption{Average value of $r^{-2}$ in $GeV^2$ of the hydrogen-like atom.
Exact Dirac (column "Exact(D)") and Schr\"{o}dinger (column "Exact(S)")
solutions are also listed for comparison to our numerical results.}
\begin{tabular}
%{|c|c|c|c|c|c|c|}
{ccccrrr}
\hline\hline

n & k & j & l & Numeric & Exact(D) & Exact(S) \\
\hline
&&&&&&\\
1 & -1 & $1\over 2$ & 0 & $1.15\times 10^{-1}$ & $1.17\times 10^{-1}$ & $6.03\times 10^{-2}$ \\
&&&&&&\\
2 & -2 & $3\over 2$ & 1 & $2.79\times 10^{-3}$ & $2.79\times 10^{-3}$& $2.51\times 10^{-3}$\\
&&&&&&\\
2 & 1 & $1\over 2$ & 1 & $5.82\times 10^{-3}$ & $5.84\times 10^{-3}$ & $2.51\times 10^{-3}$\\
&&&&&&\\
2 & -1 & $1\over 2$ & 0 & $1.83\times 10^{-2}$ & $1.87\times 10^{-2}$ & $7.54\times 10^{-3}$\\
&&&&&&\\
3 & -3 & $5\over 2$ & 2 & $1.97\times 10^{-2}$ & $1.97\times 10^{-2}$ & $1.93\times 10^{-2}$\\
&&&&&&\\
3 & 2 & $3\over 2$ & 2 & $5.14\times 10^{-4}$ & $5.14\times 10^{-4}$ & $4.47\times 10^{-4}$\\
&&&&&&\\
3 & -2 & $3\over 2$ & 1 & $8.67\times 10^{-4}$ & $8.66\times 10^{-4}$ & $7.45\times 10^{-4}$\\
&&&&&&\\
3 & 1 & $1\over 2$ & 1 & $1.79\times 10^{-3}$ & $1.80\times 10^{-3}$ & $7.45\times 10^{-4}$\\
&&&&&&\\
3 & -1 & $1\over 2$ & 0 & $5.44\times 10^{-3}$ & $5.56\times 10^{-3}$ & $2.23\times 10^{-3}$\\
&&&&&&\\
\hline\hline
\end{tabular}
\label{hydron-rn2}
\end{table}

\begin{figure}[htb]
\centering
\includegraphics[width=0.9\textwidth]{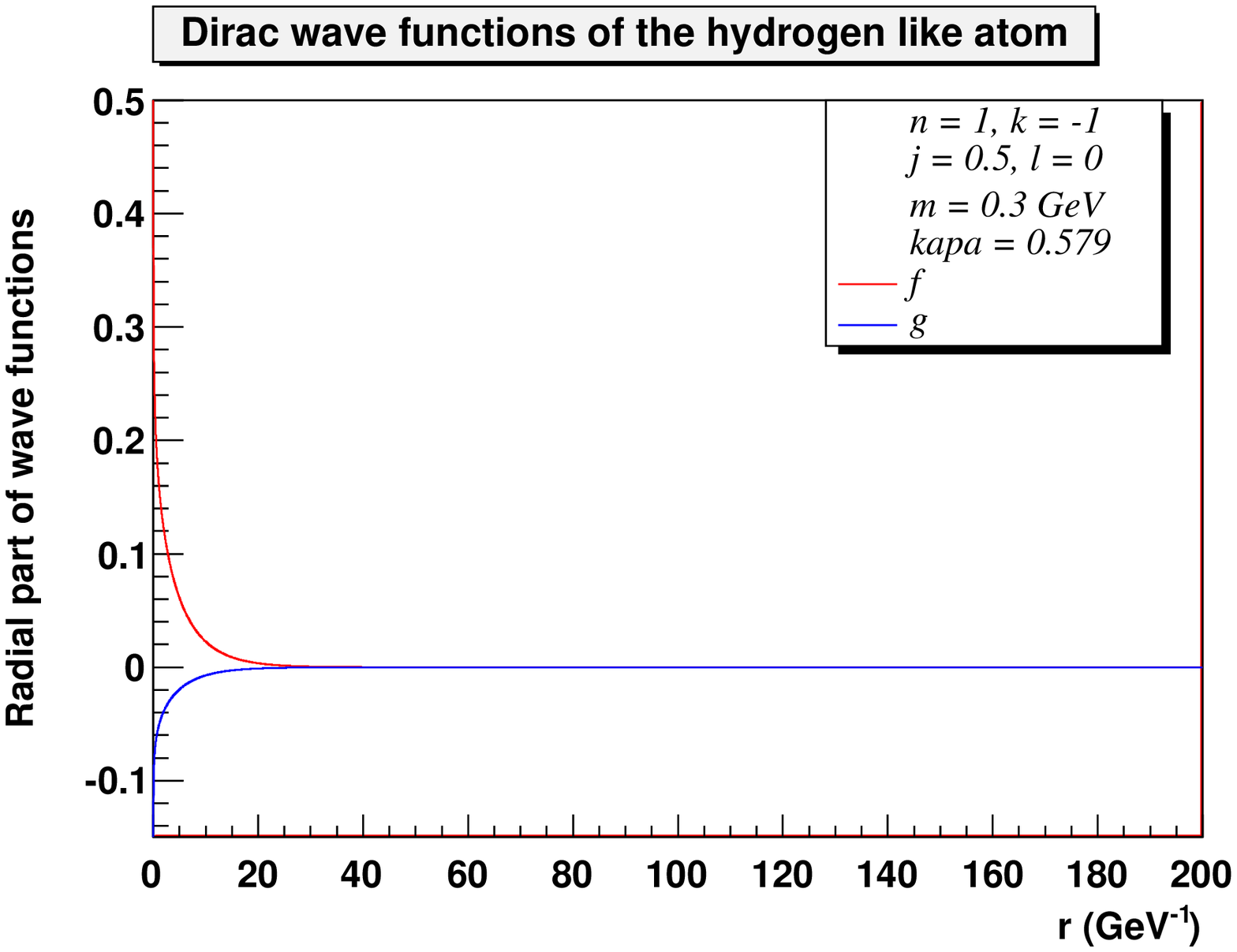}
\caption{Numerical results of the Dirac wave function of the hydrogen-like atom's $1S_{1/2}$ state.}
\label{h1-1}
\end{figure}

\begin{figure}[htb]
\centering
\includegraphics[width=0.9\textwidth]{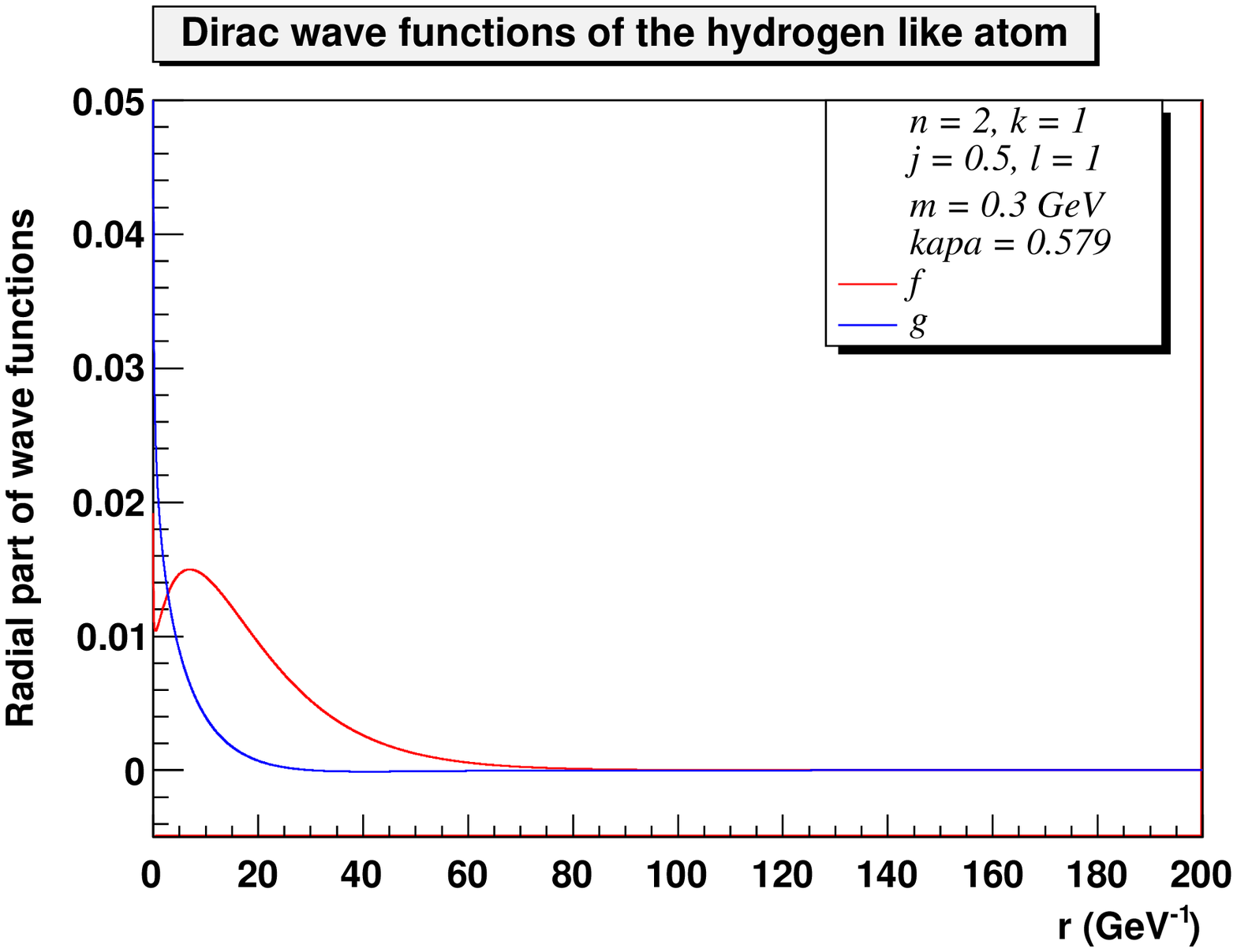}
\caption{Numerical results of the Dirac wave function of the hydrogen-like atom's $2P_{1/2}$ state.}
\label{h21}
\end{figure}

\begin{figure}[htb]
\centering
\includegraphics[width=0.9\textwidth]{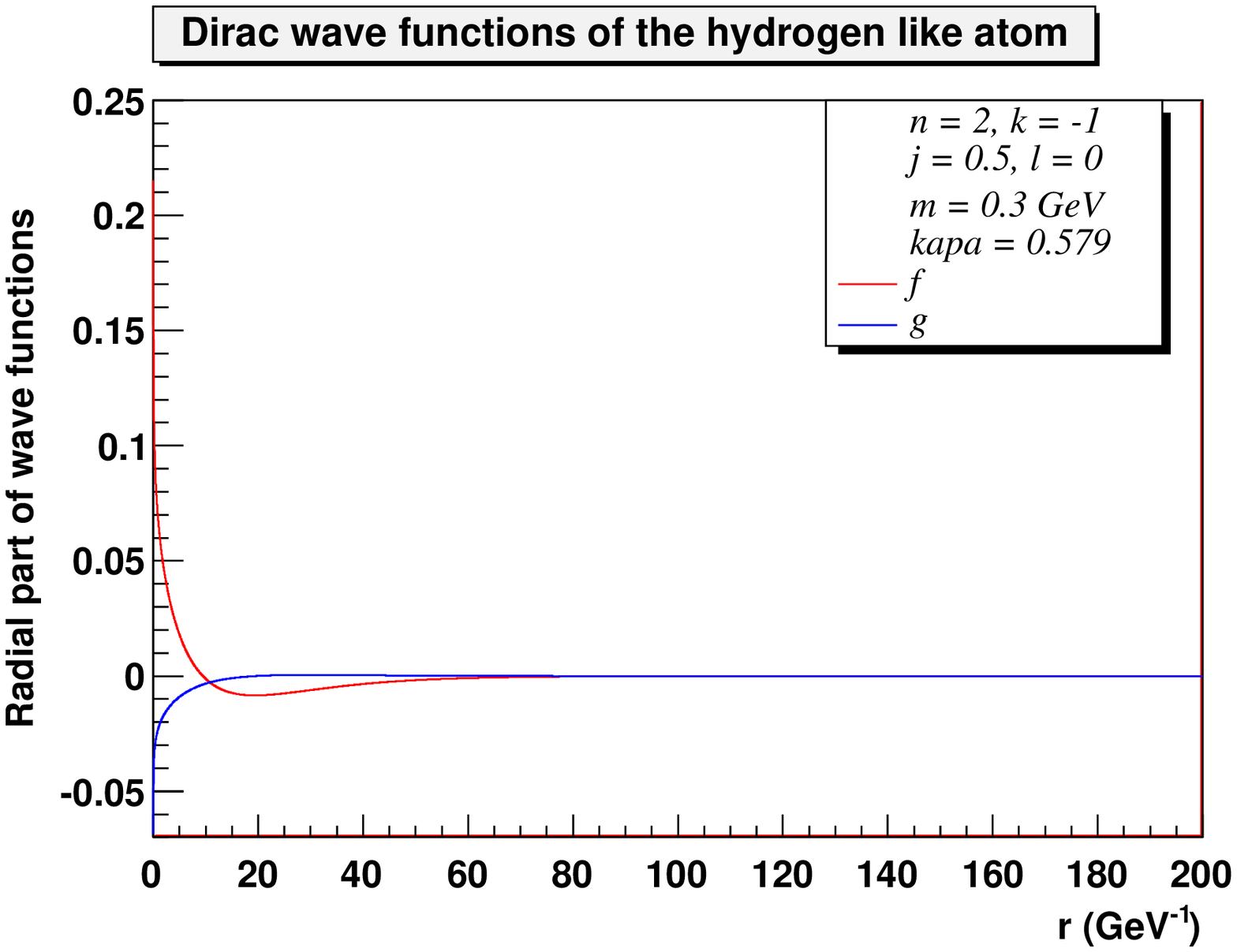}
\caption{Numerical results of the Dirac wave function of the hydrogen-like atom's $2S_{1/2}$ state.}
\label{h2-1}
\end{figure}

\begin{figure}[htb]
\centering
\includegraphics[width=0.9\textwidth]{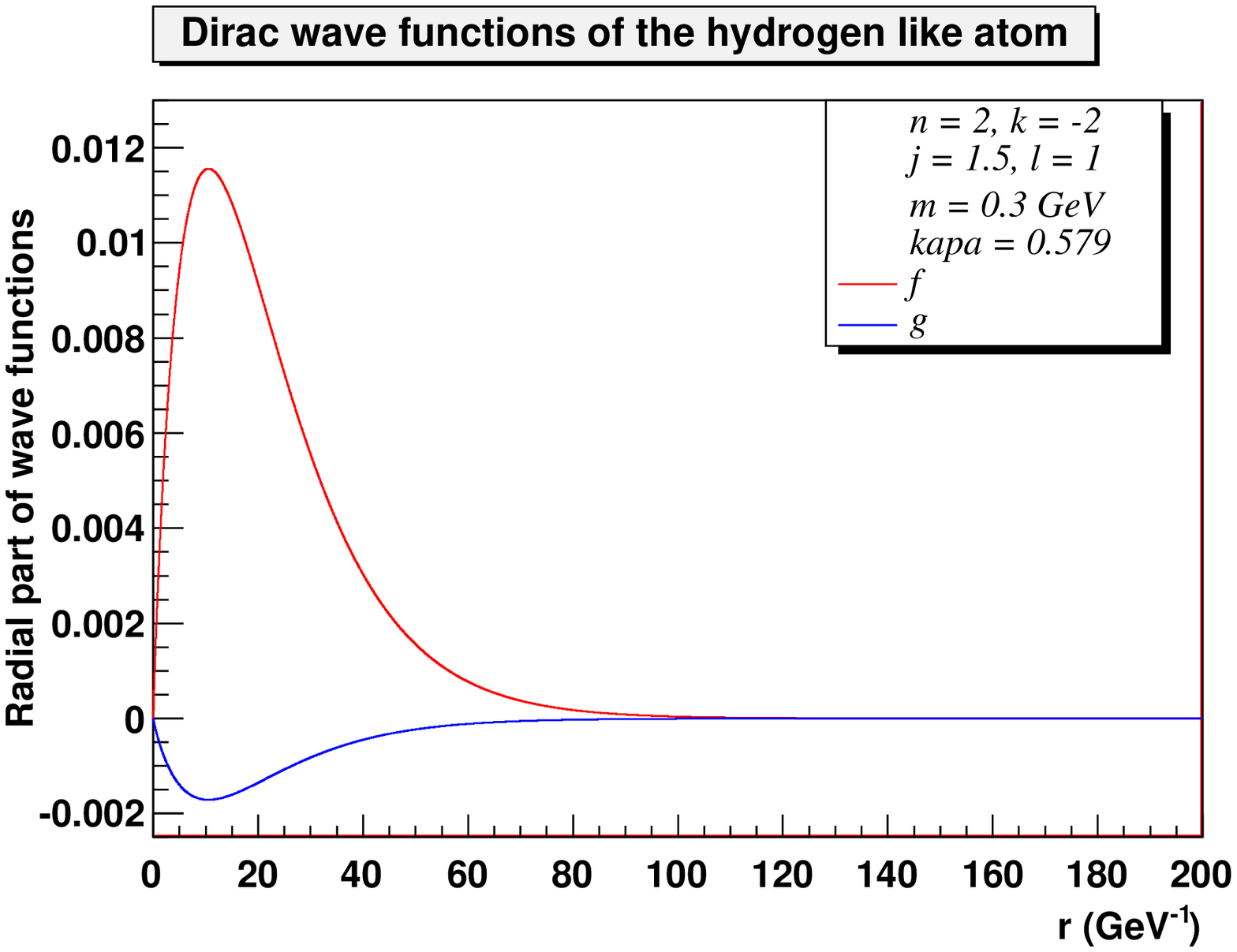}
\caption{Numerical results of the Dirac wave function of the hydrogen-like atom's $2P_{3/2}$ state.}
\label{h2-2}
\end{figure}

\begin{figure}[htb]
\centering
\includegraphics[width=0.9\textwidth]{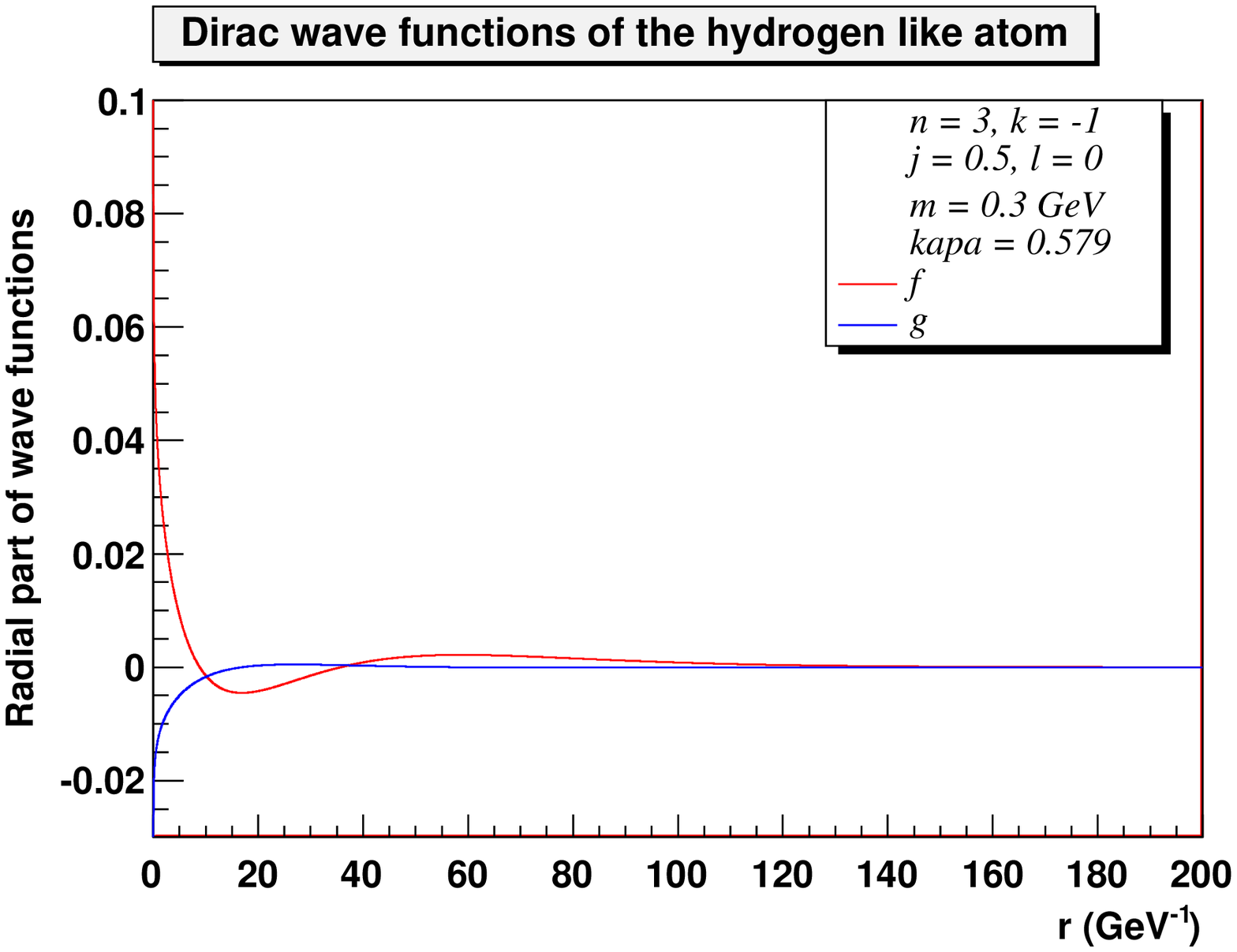}
\caption{Numerical results of the Dirac wave function of the hydrogen-like atom's $3S_{1/2}$ state.}
\label{h3-1}
\end{figure}

\begin{figure}[htb]
\centering
\includegraphics[width=0.9\textwidth]{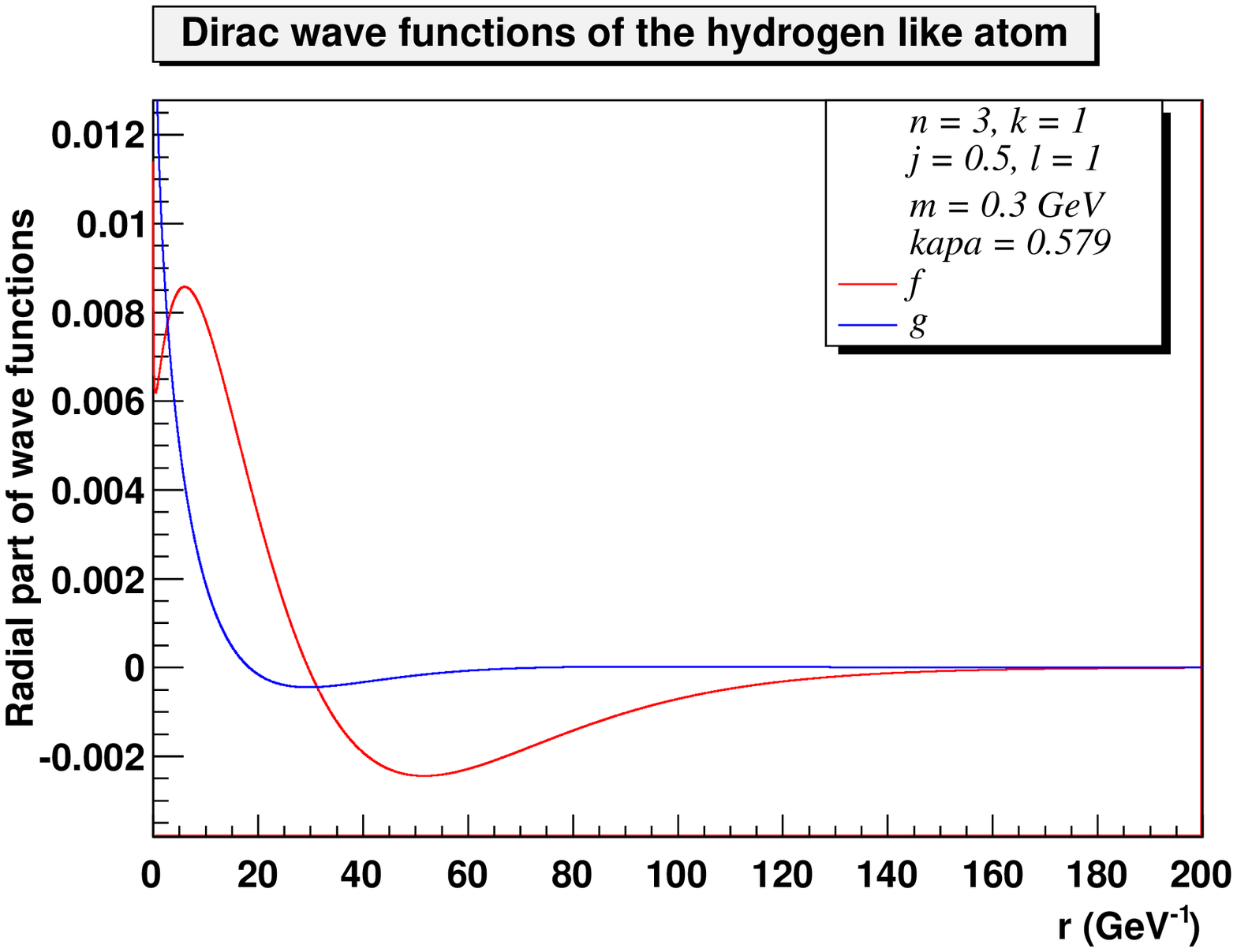}
\caption{Numerical results of the Dirac wave function of the hydrogen-like atom's $3P_{1/2}$ state.}
\label{h31}
\end{figure}

\begin{figure}[htb]
\centering
\includegraphics[width=0.9\textwidth]{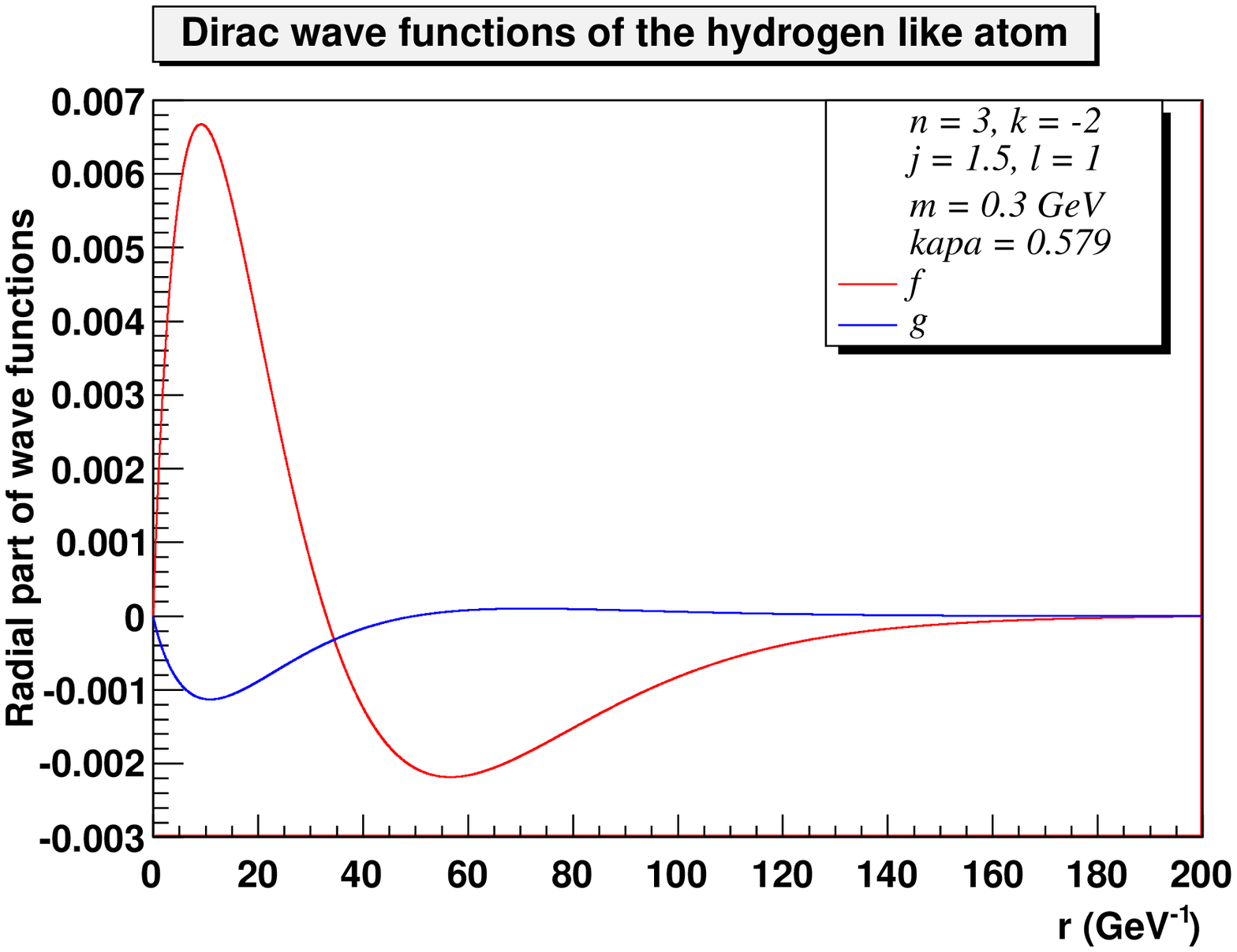}
\caption{Numerical results of the Dirac wave function of the hydrogen-like atom's $3P_{3/2}$ state.}
\label{h3-2}
\end{figure}

\begin{figure}[htb]
\centering
\includegraphics[width=0.9\textwidth]{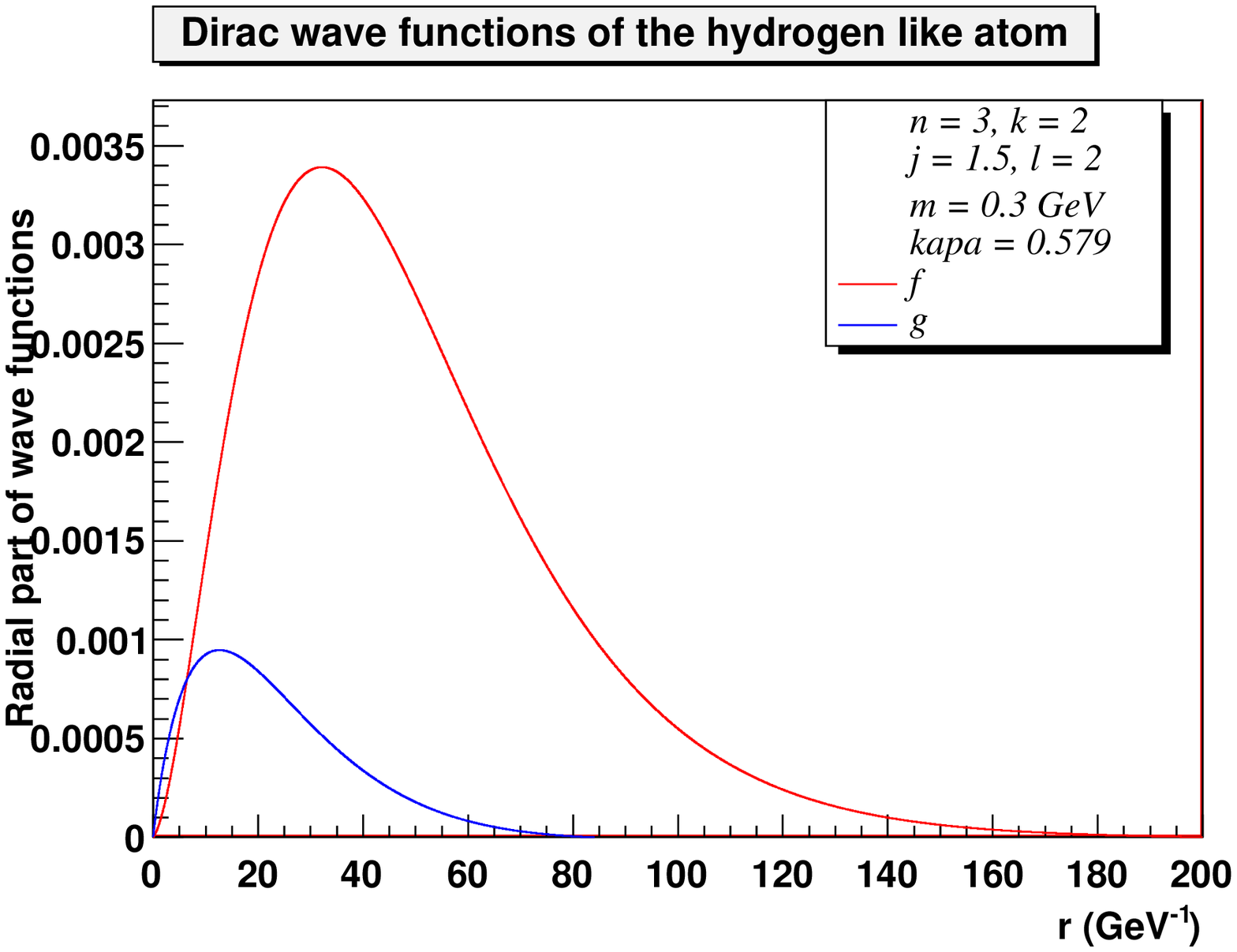}
\caption{Numerical results of the Dirac wave function of the hydrogen-like atom's $3D_{3/2}$ state.}
\label{h32}
\end{figure}

\begin{figure}[htb]
\centering
\includegraphics[width=0.9\textwidth]{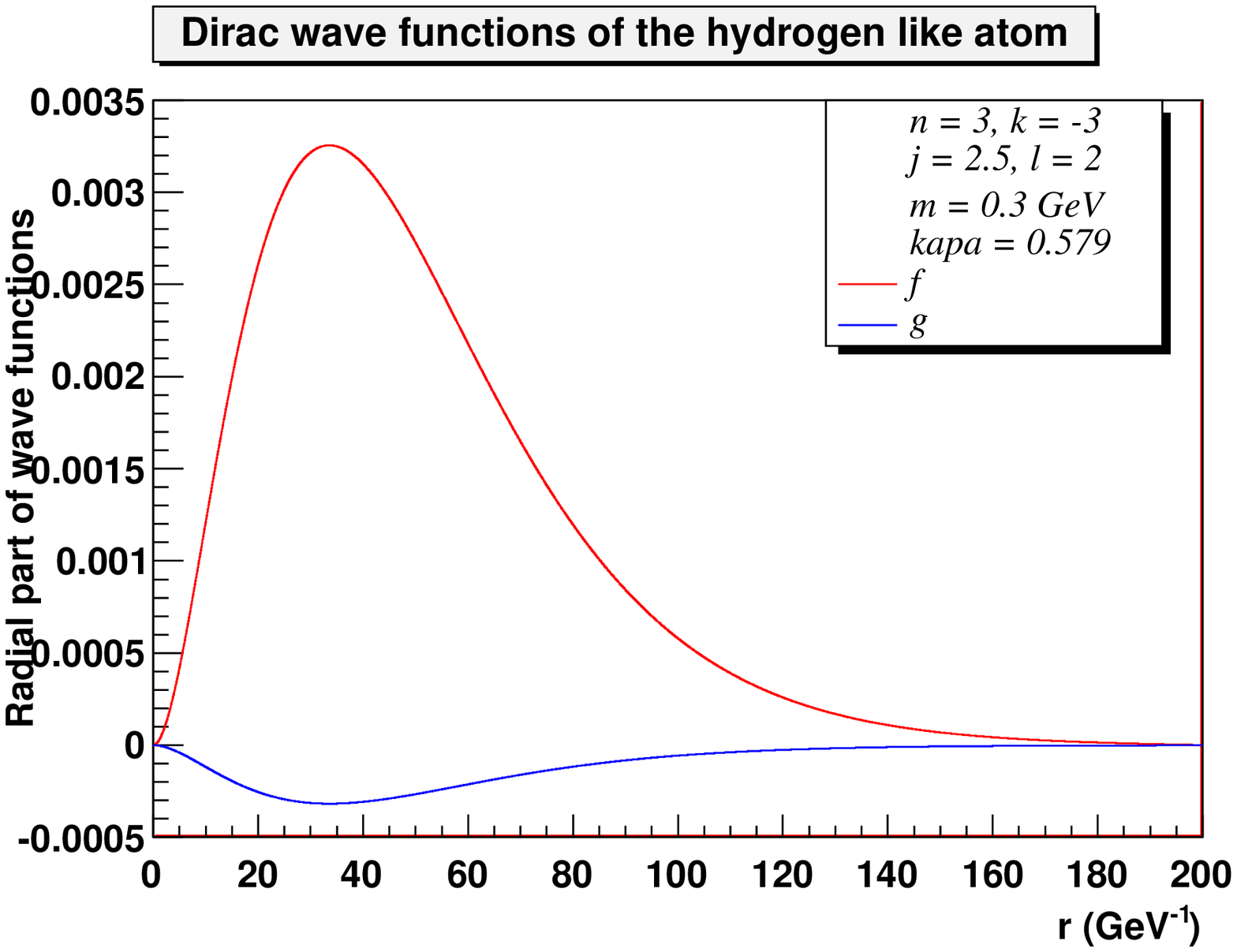}
\caption{Numerical results of the Dirac wave function of the hydrogen-like atom's $3D_{5/2}$ state.}
\label{h3-3}
\end{figure}

%========================================================
%\afterpage{\clearpage}

\section*{Appendix B: Numerical Results of with Olsson's Parameters}

The parameters in Olsson's paper~\cite{Olsson} are
\begin{eqnarray}
m_{ud} &=& 0.300 \, GeV\nonumber\\
m_s &=& 0.463 \, GeV\nonumber\\
m_c &=& 1.301 \, GeV\nonumber\\
m_b &=& 4.639 \, GeV\nonumber\\
a &=& 0.308 \, GeV^2\nonumber\\
\kappa &=& 0.579,
\label{olsson-para}
\end{eqnarray}
with the potential
\begin{eqnarray}
V_s(r) &=& ar, \\
V_v(r) &=& - \frac {\kappa}{r}.
\end{eqnarray}

\begin{table}[htb]
\centering
\caption{Using Dirac equation to fit the meson spectrum with Olsson's
parameters (\ref{olsson-para})~\cite{Olsson}. The values of mass are in $MeV$.}
\begin{tabular}
{ccccccc}
\hline\hline
 & & Spin & & Numerical & Numerical & \\
States & ${}^{2S+1}L_J$ & averaged  & $k$ & center & splitting & Parameter \\
 & & mass & & mass & mass & b \\
\hline
$\matrix{{D(1867)}\\{D^*(2010)}}$ &$\matrix{{{}^{1}S_0}\\{{}^{3}S_1}}$ & $1S(1974)$ & -1 & 1975 & $\matrix{{1867}\\{2010}}$ & b=1.64\\
%\hline
&&&&&&\\
$\matrix{{D_{1}(2423)}\\{D^{*}_2(2457)}}$ &$\matrix{{{}^{1}P_1}\\{{}^{3}P_2}}$ & $2P(2444)$ & -2 & 2444 & $\matrix{{2535}\\{2428}}$ & N/A\\
%\hline
&&&&&&\\
$\matrix{{D_{s}(1969)}\\{D^{*}_s(2110)}}$ &$\matrix{{{}^{1}S_0}\\{{}^{3}S_1}}$ & $1S(2075)$ & -1 & 2074 & $\matrix{{1968}\\{2109}}$ & b=1.61\\
&&&&&&\\
$\matrix{{D_{s1}(2535)}\\{D_{sJ}(2573)}}$ &$\matrix{{{}^{1}P_1}\\{{}^{3}P_2}}$ & $2P(2559)$ & -2 & 2559 & $\matrix{{2537}\\{2660}}$ & N/A\\
&&&&&&\\
$\matrix{{B(5279)}\\{B^{*}(5325)}}$ &$\matrix{{{}^{1}S_0}\\{{}^{3}S_1}}$ & $1S(5314)$ & -1 & 5314 & $\matrix{{5279}\\{5325}}$ & b=1.43\\
&&&&&&\\
$\matrix{{B_{s}(5375)}\\{B^{*}_s(5421)}}$ &$\matrix{{{}^{1}S_0}\\{{}^{3}S_1}}$ & $1S(5410)$ & -1 & 5412 & $\matrix{{5376}\\{5422}}$ & b=1.43\\
\hline\hline
\end{tabular}
\label{olsson-result}
\end{table}

We solve the Dirac equation with Olsson's parameters(\ref{olsson-para}).
Our fitting results of the meson's mass spectrum for both center average mass
and energy splitting are listed in Table~\ref{olsson-result} with Olsson's parameters.
The results show that we can reproduce the spin average center mass values,
which are listed in Osson's paper. Because we have the parameter $b$,
which is the width of the $\delta$ function,
we can get good results for the fine splitting of the $S$ states
by adjusting the values of $b$. On the other hand, the calculated values
of the splitting for the $P$ states are unfortunately not so good,
with the errors are around 90 MeV.

\begin{table}[htb]
\caption{Average value of $r$, $r^2$, $r^{-1}$ and $r^{-2}$,
with values are in $GeV^{n}$, by using Olsson's parameters.}
\centering
\begin{tabular}
%{|c|c|c|c|c|c|c|}
{ccrrrr}
\hline\hline
&&&&&\\
Light quark & ${}^{2S+1}L_J$ & $<r>$  & $<r^2>$ & $<r^{-1}>$ & $<r^{-2}>$ \\
\hline

u/d & ${\begin{array}{*{20}c} {}^{1}S_0 \\ {}^{3}S_1 \\ \end{array}}$ & 1.482 & 2.736 & 0.956 & 1.899\\
%\hline
&&&&&\\

u/d &
${\begin{array}{*{20}c} {}^{1}P_1 \\ {}^{3}P_2 \\ \end{array}}$ & 2.286 & 5.867 & 0.507 & 0.313\\
&&&&&\\

s & ${\begin{array}{*{20}c} {}^{1}S_0 \\ {}^{3}S_1 \\ \end{array}}$ & 1.369 & 2.353 & 1.046 & 2.306\\
&&&&&\\

s & ${\begin{array}{*{20}c} {}^{1}P_1 \\ {}^{3}P_2 \\ \end{array}}$ & 2.177 & 5.333 & 0.534 & 0.349\\
\hline\hline
\end{tabular}
\label{Olsson-expect}
\end{table}

\begin{figure}[htb]
\centering
\includegraphics[width=0.9\textwidth]{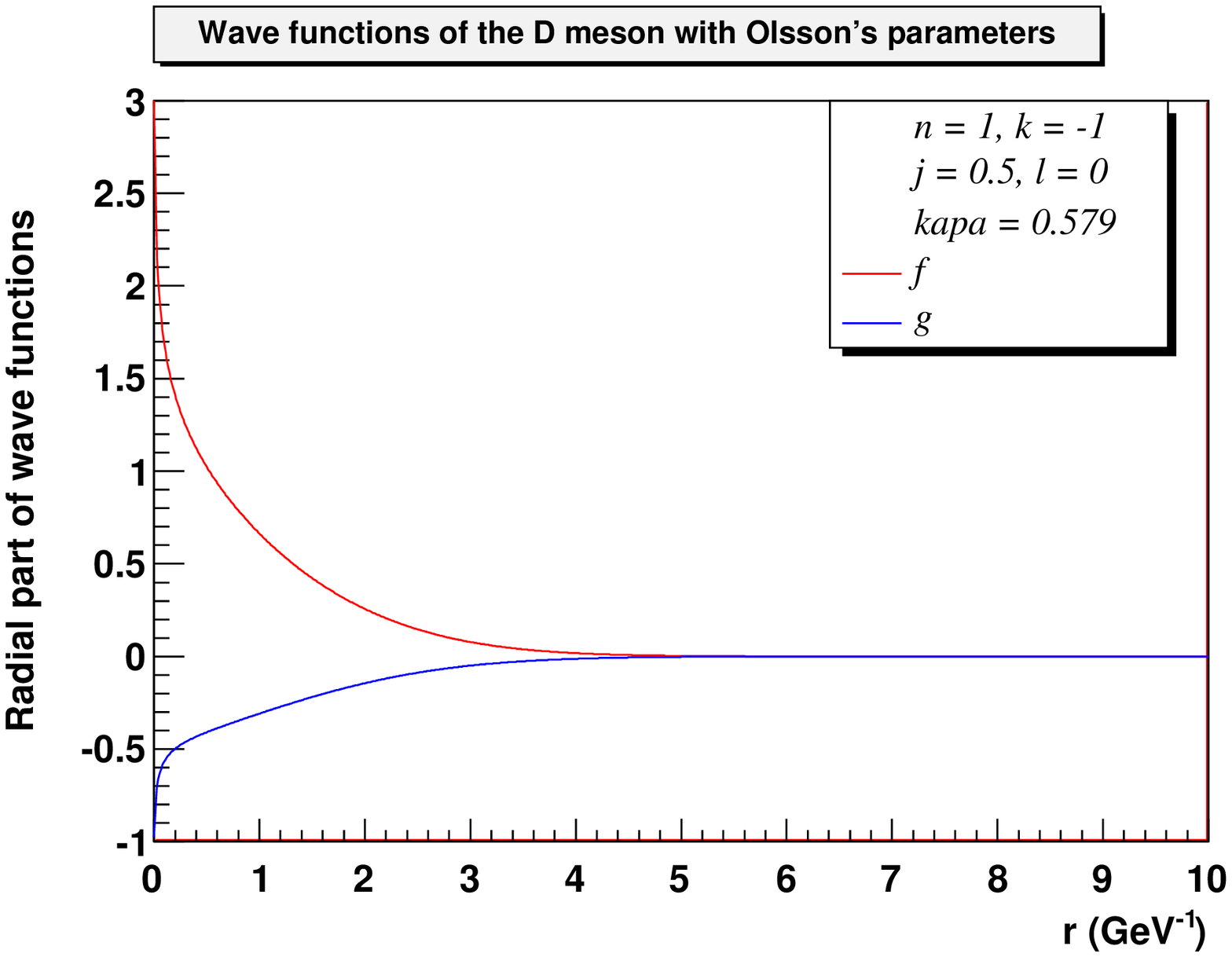}
\caption{Wave functions of the $1S$ state with Olsson's parameters.}
\label{O1}
\end{figure}

\begin{figure}[htb]
\centering
\includegraphics[width=0.9\textwidth]{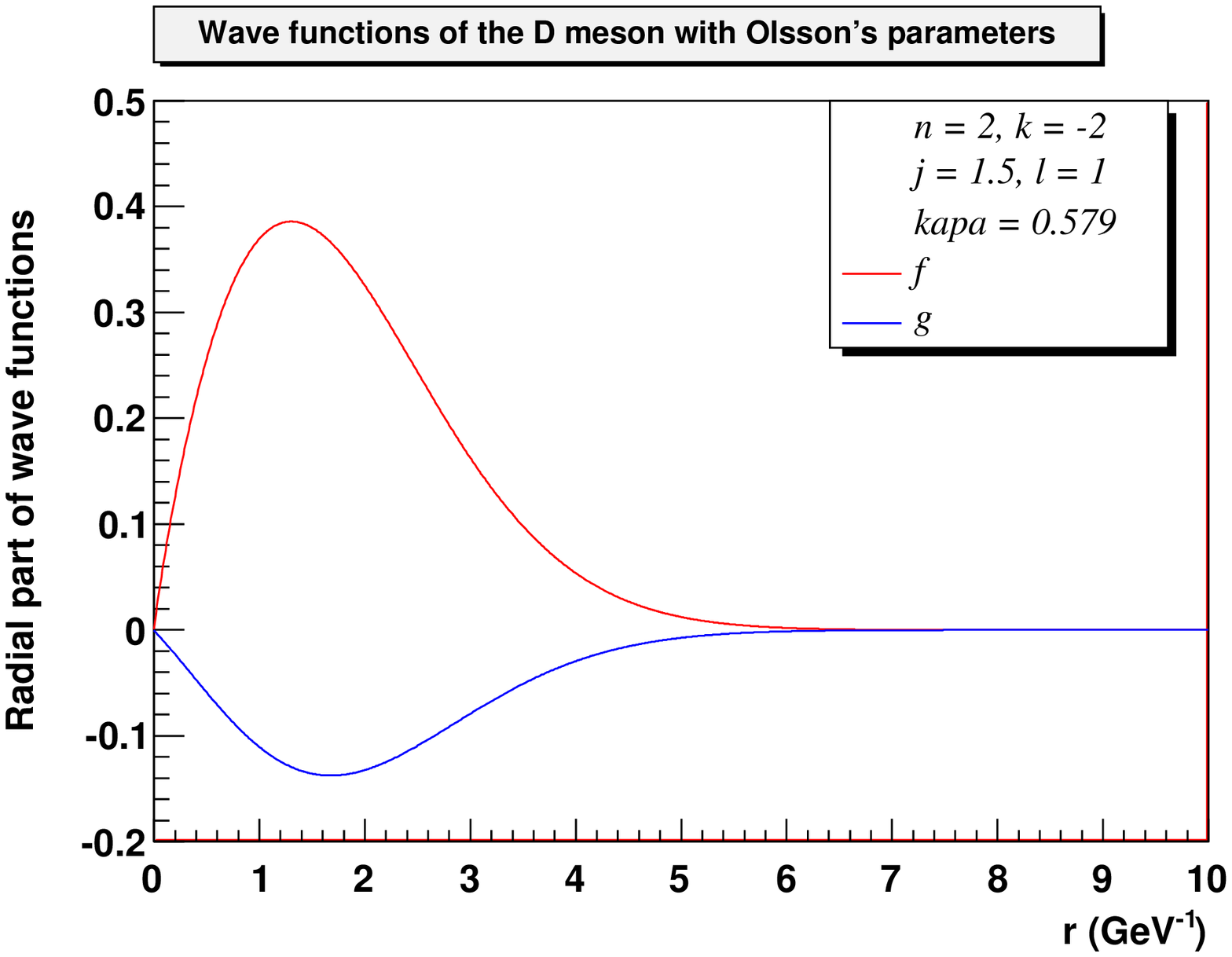}
\caption{Wave functions of the $2P$ state with Olsson's parameters.}
\label{O2}
\end{figure}

\end{document}